\documentclass[sigconf,screen]{acmart}

\usepackage{xspace}
\usepackage{color, colortbl}  

\usepackage{adjustbox}
\usepackage{siunitx}
\usepackage{tabularx} 
\usepackage{amsmath}  

\usepackage{xcolor}
\definecolor{dgrey}{gray}{0.2}

\definecolor{ao(english)}{rgb}{0.0, 0.5, 0.0}

\definecolor{violet}{HTML}{6a51a3}

\usepackage[english]{babel}  
\addto\extrasenglish{

}

\usepackage[capitalize]{cleveref} 
\usepackage[inline]{enumitem} 

\newcommand{\myparagraph}[1]{\paragraph*{\hspace*{-\parindent}\normalsize\bf#1}}
\AtBeginDocument{%
  }

\copyrightyear{2025}
\acmYear{2025}
\setcopyright{cc}
\setcctype{by-nd}
\acmConference[SIGIR '25]{Proceedings of the 48th International ACM SIGIR Conference on Research and Development in Information Retrieval}{July 13--18, 2025}{Padua, Italy}
\acmBooktitle{Proceedings of the 48th International ACM SIGIR Conference on Research and Development in Information Retrieval (SIGIR '25), July 13--18, 2025, Padua, Italy}\acmDOI{10.1145/3726302.3730174}
\acmISBN{979-8-4007-1592-1/2025/07}

\begin{document}

\title[Characterising Topic Familiarity and Query Specificity Using Eye-Tracking Data]{Characterising Topic Familiarity and Query Specificity\\Using Eye-Tracking Data}

\author{Jiaman He}
\orcid{0009-0007-2817-7675}
\affiliation{%
\institution{RMIT University}
\city{Melbourne}
\country{Australia}
}
\email{jiaman.he@student.rmit.edu.au}

\author{Zikang Leng}
\orcid{0000-0001-6789-4780}
\affiliation{%
\institution{Georgia Institute of Technology}
\city{Atlanta}
\country{USA}
}
\email{zleng7@gatech.edu}

\author{Dana McKay}
\orcid{0000-0001-7522-1842}
\affiliation{%
\institution{RMIT University}
\city{Melbourne}
\country{Australia}
}
\email{dana.mckay@rmit.edu.au}

\author{Johanne R. Trippas}
\orcid{0000-0002-7801-0239}
\affiliation{%
\institution{RMIT University}
\city{Melbourne}
\country{Australia}
}
\email{j.trippas@rmit.edu.au}

\author{Damiano Spina}
\orcid{0000-0001-9913-433X}
\affiliation{%
\institution{RMIT University}
\city{Melbourne}
\country{Australia}
}
\email{damiano.spina@rmit.edu.au}

\renewcommand{\shortauthors}{Jiaman He, Zikang Leng, Dana McKay, Johanne R. Trippas, and Damiano Spina}

\begin{abstract}

Eye-tracking data has been shown to correlate with a user's knowledge level and query formulation behaviour. While previous work has focused primarily on eye gaze fixations for attention analysis, often requiring additional contextual information, our study investigates the memory-related cognitive dimension by relying solely on pupil dilation and gaze velocity to infer \emph{users' topic familiarity} and \emph{query specificity} without needing any contextual information. Using eye-tracking data collected via a lab user study ($N=18$), we achieved a Macro F1 score of 71.25\% for predicting topic familiarity with a Gradient Boosting classifier, and a Macro F1 score of 60.54\% with a k-nearest neighbours (KNN) classifier for query specificity. Furthermore, we developed a novel annotation guideline -- specifically tailored for question answering -- to manually classify queries as Specific or Non-specific. This study demonstrates the feasibility of eye-tracking to better understand topic familiarity and query specificity in search.

\end{abstract}



\maketitle

\section{Introduction}
\label{sec:intro}
When people search for information, they are typically driven by their realisation of a knowledge gap~\cite{belkin1980anomalous}, which motivates them to recall relevant prior knowledge. Search begins when individuals feel motivated to address that gap. As motivation builds, they start looking for information to bridge that gap, which involves a cognitive process of integrating new with old knowledge.

A user's familiarity with a topic has been defined as the extent of the user's prior knowledge~\cite{chesky1987effects}. 
Understanding this familiarity allows search systems to tailor results to the searcher's knowledge level, which is also linked to the query formulation behaviour\cite{eickhoff2015eye}. Query specificity is a critical factor in interpreting user intent~\cite{gonzalez2011web}.

By analysing both \textit{topic familiarity} and \textit{query specificity} through physical responses, search systems could adapt to users' needs in real-time. Pupil responses and eye movements reflect cognitive process like memory and recognition~\cite{goldinger2012pupil}. With mobile and wearable devices (e.g., Tobii Pro Glasses 3~\cite{tobiiProGlasses3}, Apple Vision Pro~\cite{appleVisionPro}) now supporting eye-tracking~\cite{krafka2016eye}, such data can be captured in real-world scenarios.

While eye-tracking data has been effectively used to characterise how people pay attention to search results in search engine results pages (SERPs)~\cite{miller2005attention,buscher2012attentive,palani2020eye,dumais2010individual,mao2017understading,harris2019detecting}, using eye-trackers to characterise the realisation of an information need and the query formulation in search processes remain under-explored~\cite{ji2024characterizing}. 

This work aims to characterise \textit{topic familiarity} and \textit{query specificity} using only eye-tracking data, which could inform the development of an adaptive information retrieval (IR) system based on how people look at the screen.
\noindent Our contributions are as~follows:

\begin{enumerate}[topsep=0pt]

    \item It is feasible to classify users’ knowledge level in real time using pupil dilation and gaze velocity, without relying on additional contextual information, as demonstrated by our \textit{topic familiarity} prediction (Macro F1 = 71.25\% with a Gradient Boosting classifier). 

    \item We developed a novel methodology to annotate the specificity of users' search queries. Using this method, we categorised a set of 83 queries as either Specific or Non-specific.\footnote{The annotated dataset is publicly available at \url{https://github.com/peanutH/Familiarity-QuerySpec}, and includes the queries and their specificity classifications} 

    \item It is feasible to classify query formulation behaviour using pupillary data alone as demonstrated by our \textit{query specificity} classification (Specific vs. Non-specific), which achieved a Macro F1 score of 60.54\% with a KNN classifier.

\end{enumerate}

\section{Related Work}
\label{sec:relatedwork}

\myparagraph{Topic Familiarity in Information Seeking}

Information seeking begins with an understanding that there is something we do not know (an Anomalous State of Knowledge, ASK) and concludes with that knowledge gap being resolved~\cite{belkin1980anomalous}. This process includes integrating prior knowledge and searching for new information, which may reveal other knowledge gaps. Information seeking is thus a cyclical, iterative process, as described by~\citet{marchionini1995information}.
When people notice an information gap, they review their existing knowledge; if their knowledge falls short, they seek additional information.
This cycle repeats until the knowledge gap is resolved, whereupon information-seeking stops. 

Aligned with this search cycle is the dual-process cognitive model, which posits that memory involves two distinct mechanisms: \textit{recollection} and \textit{familiarity}~\cite{atkinson1972search}. 
Recollection involves retrieving details about past events, while familiarity enables individuals to distinguish between previously encountered and new information~\cite{yonelinas2001components}. Together, these processes help people recognise familiar topics and identify novel ones. Research has shown that such memory-related processes can be inferred from eye movement patterns~\cite{ryan2000amnesia}. 
Prior work has also demonstrated that a user's domain knowledge can be inferred from their eye movements~\cite{cole2013inferring}, suggesting that search tools could leverage real-time eye-tracking data to estimate user knowledge, allowing for more personalised search experiences.

\myparagraph{Query Specificity.}
To form a sentence about an idea, including constructing a query, we first need at least some understanding of the idea~\cite{wundt1970psychology}. Query formulation thus occurs after cognitive memory retrieval and activation of prior knowledge. As topic familiarity, defined as the extent of prior knowledge~\cite{chesky1987effects}, has been shown to influence query formulation behaviour~\cite{hu2013effects}, query specificity offers an opportunity to interpret user intent~\cite{gonzalez2011web}. 

Early work in this space focused on keyword queries, but searchers increasingly use natural language (NL) queries, leading to more diverse query structures and intentions. To address the evolving query forms, we developed an innovative annotation method grounded in NL questions and answers to classify query specificity.

In information-seeking tasks, the meaning of a question lies in its set of possible answers~\cite{rooy2004utility}. According to \citet{hamblin1976questions}, understanding a question's meaning requires knowing what qualifies as an appropriate answer. 
We formulated guidelines for annotating queries as Specific and Non-specific. The methodology for query specificity classification is discussed in~\autoref{subsec:dataset}. 

\myparagraph{Eye Features.} 
Previous work has explored eye fixations with query suggestions~\cite{eickhoff2015eye} and the inference of domain knowledge~\cite{cole2013inferring}. Our work builds on the idea that topic familiarity and query formulation are memory-related cognitive processes. We investigated eye features associated with human memory. In particular, pupil size has been shown to effectively isolate cognitive effects of familiarity from physical stimulus characteristics~\cite{franzen2022individual}.  
Furthermore, research has shown distinct pupillary responses based on familiarity with branded products, highlighting the link between pupil dynamics and recognition processes~\cite{franzen2022individual}. In addition to pupil size, eye gaze has been used to automatically detect internal states of familiarity~\cite{castillon2024automatically}, as eye movement patterns are closely linked to cognitive memory recall~\cite{bulling2011recognition}.
\section{Experimental Evaluation}
\label{sec:Experiment}
\begin{figure}[tp]
    \centering
    \includegraphics[width=\linewidth]{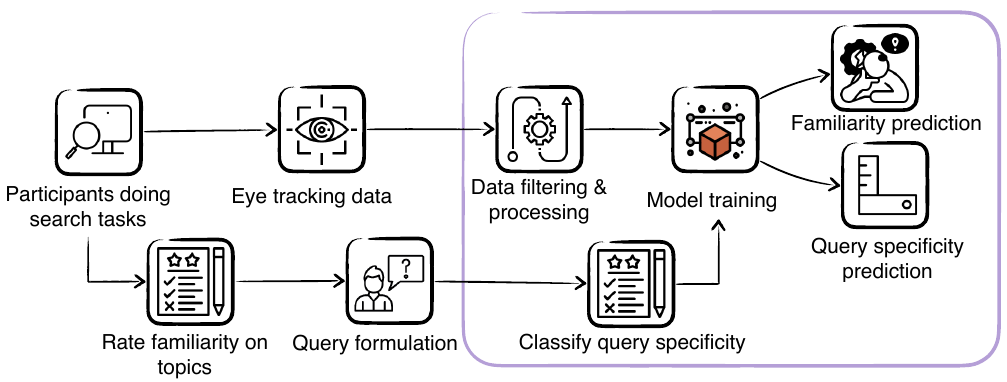}
    \caption{Flow of the experiment; our work enclosed in box. 
    }
    \label{fig:pipeline}
\end{figure}

Participants performed a search task, during which eye-tracking data, topic familiarity ratings, and query specificity were collected (\autoref{subsec:dataset}). 
\autoref{fig:pipeline} illustrates the experimental pipeline. 
The data was preprocessed and cleaned (\autoref{subsec:data preprocessing}) before training five classifiers to predict \textit{topic familiarity} and \textit{query specificity} (\autoref{subsec:lassifier training}). 
\vspace{-0.15in}

\subsection{Dataset}
\label{subsec:dataset}
\begin{figure}[tp]
    \centering
    \includegraphics[width=\linewidth]{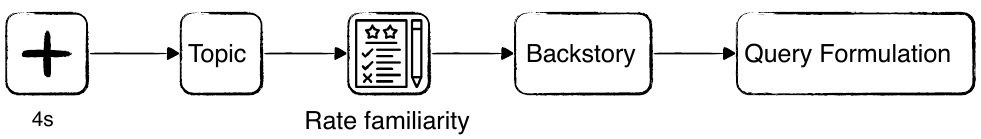}
    \caption{Experiment overview by~\citet{ji2024characterizing}.}  
    \label{fig:experiment}
\end{figure}

We used user study data collected by~\citet{ji2024characterizing}. In the study, participants focused on a cross in the centre of a blank screen for 4 seconds. Then, one of the 12 topic titles appeared, and participants rated the familiarity on a 5-point scale. Next, a backstory was provided for the query formulation, where the backstory was selected from the \textit{InformationNeeds} dataset~\cite{bailey2015information} to evoke the users' realisation of the knowledge gap or information need.
The Tobii Pro Fusion eye-tracker\footnote{\textcolor{blue}{\href{https://www.tobii.com/products/eye-trackers/screen-based/tobii-pro-fusion}{https://www.tobii.com/products/eye-trackers/screen-based/tobii-pro-fusion}}} was used to collect eye tracking data (60Hz). \autoref{fig:experiment} shows an overview of the procedure.

To ensure the dataset's consistency, we use only written queries in the experiment by~\citet{ji2024characterizing}, as voice input influences query characteristics such as length, terminology, and language~\cite{guy2018characteristics, vtyurina2020mixed, trippas_towards_2020}. We then analyse the eye-tracking data collected during the phases before the backstory phase, including pupil size and eye~gaze.

We re-scaled \textit{familiarity}, measured on a 5-point to binary scale (\textit{1} as unfamiliar and \textit{2} familiar). This approach aligns with familiarity as a signal detection process, where individuals assess whether something feels familiar or unfamiliar~\cite{yonelinas2001components}. Specifically, 1--3 ratings were grouped as 1 (unfamiliar), while 4--5 as 2 (familiar). We classified a rating of 3 as unfamiliar, based on the assumption that participants selecting this midpoint were uncertain about their familiarity, which we interpreted as indicative of a lack of familiarity.\footnote{We have additionally experimented with classifying the rating of 3 as familiar and observed similar model performance trend with overall lower performance.}

To annotate query specificity, \citet{hafernik2013understanding} showed that queries can be classified as \textit{Specific} or \textit{Non-specific} using a list of attributes. However, their work, conducted over a decade ago, focused solely on keyword-based queries. In contrast, our work examines mostly NL queries, and when three authors applied these attributes to our dataset, the resulting annotator agreement was low. Next, we developed detailed query specificity annotation~guidelines:

\begin{enumerate}[label=(\roman*), font=\itshape]
    \item Does the query have a clear objective answer? (Yes $\rightarrow$ jump to \textit{(ii)} ; No/Not sure $\rightarrow$ Non-specific)
    \item Does the query require exhaustively listing all valid propositions? (Yes $\rightarrow$ jump to \textit{(iv)}; No/Not sure $\rightarrow$ jump to  \textit{(iii)})
    \item Is it a single/unique answer? (Yes $\rightarrow$ Specific; No/Not sure $\rightarrow$ jump to \textit{(v)})
    \item Is the answer set bounded? (Yes $\rightarrow$ Specific; No/Not sure $\rightarrow$  Non-specific)
    \item Does the answer suffice to be useful and not misleading without requiring exhaustivity? (Yes $\rightarrow$ Specific; No/Not sure $\rightarrow$ Non-specific)
\end{enumerate}

For keyword-based queries, we treated them as NL queries by reframing them as questions. For example, the keyword query ``greek plays'' is re-imagined as ``What are Greek plays?'' to align with NL query processing.

Three authors independently annotated the queries, with the final annotation determined by majority agreement. We observed a full agreement of 71$\%$ and a Fleiss' Kappa~\cite{fleiss1971measuring} of $0.56$ among annotators. \autoref{table:distribution} presents the distribution of topic familiarity and query specificity. An example of a Specific query is ``What are the sources of slate stone for decorative use, and how is it obtained?'' and an example of a Non-Specifc query is  ``What is the price of the stone?''.
After preprocessing and cleaning the data (\autoref{subsec:data preprocessing}), we retained data from 18 participants, with each contributing between 2 and 6 queries. Note that the total number of data points for topic familiarity and query specificity differs, as not every participant provided a familiarity rating for every topic.

\begin{table}[tp]
\centering
\caption{Number of instances categorized by topic familiarity 
and query specificity 
.}
\adjustbox{max width=.48\textwidth}{
\begin{tabular}{ccccc}
\toprule
\multicolumn{2}{c}{\textbf{Topic Familiarity}}  & & \multicolumn{2}{c}{\textbf{Query Specificity}} \\
\cmidrule{1-2} \cmidrule{4-5}
 Unfamiliar (49)  & Familiar (21) & &  Non-Specific (60) & Specific (23) \\
 \bottomrule
\end{tabular}
}
\label{table:distribution}
\end{table}

\subsection{Data Preprocessing and Cleaning}
\label{subsec:data preprocessing}
The baseline data is eye-tracking data from the 4-second cross phase, where participants fixated on a central cross. We first downsampled the 60 Hz eye gaze data to 30 Hz, as recommended for pupillometry studies~\cite{winn2018best}. Next, we followed the process as described by~\citet{franzen2022individual}. The baseline preprocessing and product-viewing stage pupil data included interpolation of blinks, data smoothing, subtractive baseline correction~\cite{mathot2018safe}, removal of trials with numerous missing and/or outlier samples, and frequency downsampling. The subtractive baseline correction involved subtracting the median pupil value calculated from the last 150 ms of the baseline stage, just before the transition to the topic page~\cite{mathot2018safe}.
When pupil data was missing for one eye, we imputed the missing value using the corresponding value from the other eye~\cite{blumenfeld2002neuroanatomy}.
The pupil feature was then calculated as the mean of both eyes' values.

Previous research has demonstrated that familiar product images are associated with larger average and peak pupil dilation, with a prolonged response beginning around 1400 ms post-stimulus, indicating differences in familiarity across participants~\cite{franzen2022individual}. Based on this, we analysed eye data starting at 1.5 seconds and tested at 300 ms intervals. This approach aligns with findings that real effects on pupil size, such as constriction or dilation in response to stimuli, typically require at least 220 ms to manifest~\cite{mathot2018safe}.

For the analysis, we computed the Relative Pupil Dilation (RPD), which measures the relative change in the current pupil diameter compared to a baseline value~\cite{franzen2022individual, mathot2018safe}. RPD is defined as: $RPD_t = (P_t - P_{\text{baseline}})/P_{\text{baseline}}$.
Here, $P_t$ represents the pupil dilation at time $t$, and $P_{\text{baseline}}$ is the baseline pupil dilation value, calculated as the average pupil dilation over the 4-second cross-phase.

Additionally, we calculated the gaze velocity using the formula: $v = \sqrt{(x_2 - x_1)^2 + (y_2 - y_1)^2}/\Delta t$
where $(x, y)$ represents the coordinates of the eye gaze on the screen, and $\Delta t$ is the time difference between two successive measurements.

\subsection{Classifier Training}
\label{subsec:lassifier training} 

We use five classifiers to predict users' familiarity with a topic and query specificity based on eye-tracking data: Random Forest~\cite{breiman2001random}, KNN~\cite{fix1985discriminatory}, Logistic Regression~\cite{cox1958regression}, Gradient Boosting~\cite{Friedman2001boost}, and Decision Tree~\cite{quinlan1986tree}. Deep learning classifiers were not used, as they typically require large datasets to perform effectively.

We computed statistical and signal-based features using the extracted RPD and gaze velocity data, including mean, standard deviation, skewness, kurtosis, zero crossings, and peak-related metrics. To refine the feature set, we applied Recursive Feature Elimination with Cross-Validation (RFECV)~\cite{guyon2002gene}, which iteratively removed the least informative features. For the topic familiarity prediction, the final selected features included gaze velocity skewness, skewness, kurtosis, root mean square, and the number of peaks for RPD. For the query specificity prediction, only the number of peaks for RPD is used as the feature for the classifier.

For evaluation, we used two approaches: five-fold stratified cross-validation and leave-one-subject-out (LOSO) cross-validation~\cite{ji2023examining,ji2024characterizing}, which evaluates the models' ability to generalise to unseen users, thereby ensuring robustness in user-independent scenarios. 
For five-fold stratified cross-validation, we report the mean and standard deviation of accuracy, macro F1 score, and area under the curve (AUC) across all folds. We report the average macro F1 score across all users for LOSO cross-validation. Additionally, we conducted a paired Student's t-test to compare model performance across 18 folds (1 fold per participant) under the LOSO cross-validation setting. Specifically, we individually compared the best-performing model under LOSO cross-validation against the other four models.

\section{Results}
\label{sec:results}

\myparagraph{Predicting Topic Familiarity}
\autoref{table:familiarity} presents the performance of various models in predicting users' familiarity with a topic based on eye-tracking data. In stratified cross-validation, the Logistic Regression classifier achieved the highest accuracy (81.81\%), while the Gradient Boosting classifier achieved the highest macro F1 score (71.25\%), and the KNN classifier obtained the highest AUC (82.61\%). Under LOSO cross-validation, the Logistic Regression classifier achieved the highest average macro F1 score (63.74\%).                                             
To assess statistical significance, we conducted a paired Student’s t-test with Bonferroni correction to compare the Logistic Regression classifier’s performance in LOSO cross-validation against other models. At $\alpha/4 = 0.0125$, no statistically significant differences were found.

\myparagraph{Predicting Query Specificity}

\autoref{table:specificity} presents the results for query specificity prediction. In stratified cross-validation, the KNN classifier achieved the highest accuracy (71.24\%), and macro F1 score (60.54\%). Gradient Boosting outperformed other models in terms of AUC (74.09\%). For LOSO cross-validation, the Logistic Regression classifier demonstrated the strongest performance with an average Macro F1 score of 58.11\%.

We performed a paired Student’s t-test with Bonferroni correction to compare the Logistic Regression classifier’s performance in LOSO cross-validation against the other models. Using a significance level of $\alpha/4=0.0125$, we did not find statistically significant differences in performance.

\begin{table}[tp]
\centering
 \caption{Effectiveness ($\mu\pm\sigma$) of binary classifiers for topic familiarity using eye-tracking data.
 } 
\adjustbox{max width=.48\textwidth}{
\sisetup{
    separate-uncertainty = true,
    table-align-uncertainty = true,
    detect-weight = true
}
\begin{tabular}{lS[table-format=2.2(2)]S[table-format=2.2(2)]S[table-format=2.2(2)]S[table-format=2.2]}  
\toprule
\textbf{Model} & \textbf{Accuracy} & \textbf{Macro F1} & \textbf{AUC} & \textbf{LOSO F1} \\
\midrule

\text{Logistic Regression} & 
\cellcolor[HTML]{caebc0} \bfseries 81.81 \pm 7.34 & 
70.06 \pm 16.88 & 
76.27 \pm 5.18 & 
\cellcolor[HTML]{caebc0} \bfseries 63.74 \\

\text{KNN}       & 
77.52 \pm 10.49 & 
71.17 \pm 15.65 & 
\cellcolor[HTML]{caebc0} \bfseries 82.61 \pm 12.84 & 
50.76 \\ 

\text{Gradient Boosting}   & 
77.81 \pm 2.31 & 
\cellcolor[HTML]{caebc0} \bfseries 71.25 \pm 3.06 & 
68.64 \pm 8.68 & 
55.21 \\

\text{Random Forest}   & 
69.52 \pm 4.82 & 
57.47 \pm 5.13 & 
70.48 \pm 14.76 & 
49.69 \\

\text{Decision Tree}       & 
72.29 \pm 3.84 & 
66.39 \pm 5.85 & 
68.55 \pm 6.71 & 
51.71 \\

\bottomrule
\end{tabular}
}
\label{table:familiarity}
\end{table}

\section{Discussion}
\label{sec:discussion}

Our result shows the feasibility of using eye-tracking data as a real-time cognitive measurement to predict users' topic familiarity and query specificity. However, we observed a general trend of lower model performance under LOSO cross-validation compared to stratified cross-validation. This discrepancy can indicate that the prediction is more accurate when the model has seen some of a user’s eye-tracking data before. Because each person has unique eye movement patterns~\cite{peterson2013individual}, the model benefits from learning these individual differences, leading to better accuracy when it has access to some of a user's eye-tracking data during training.

For query specificity classification, we used only pupil dilation features. Regarding topic familiarity, four out of five features were also derived from pupil dilation, with only one related to gaze velocity. This suggests that pupillary responses may serve as a stronger indicator than eye gaze in classifying these tasks, likely due to their close relationship with cognitive workload~\cite{weber2021assessing}.
Additionally, all five models are poorer at predicting query specificity than topic familiarity. This disparity can be attributed to the task timing and the differing cognitive demands they impose. Familiarity ratings were collected after participants read about the topic, aligning closely with the eye-tracking data gathered during the reading phase. This temporal proximity ensured that participants' cognitive state during familiarity rating was consistent with the eye-tracking data. 

In contrast, query formulation occurred after familiarity rating and an additional stage involving reading a backstory (\autoref{fig:experiment}). These intermediate steps likely altered the participants' cognitive state, making the eye-tracking data collected during the initial topic-reading phase less representative of the mental processes involved in query formulation. 
Moreover, query formulation is cognitively demanding. It requires translating internal knowledge into structured language, engaging both recognition and recollection-based memory retrieval~\cite{wundt1970psychology}. This dual memory involvement increases cognitive load and variability, making accurate predictions more difficult.

Topic familiarity, on the other hand, primarily involves recognition—a process associated with lower cognitive effort, as described by the single-detection process model~\cite{atkinson1972search}. This makes familiarity easier to infer from eye-tracking data, helping to explain the superior model performance in this task.

\section{Conclusions} 
\label{sec:conclusion}
We explored the potential of eye-tracking data to  predict users' topic familiarity and query specificity as a cognitive memory retrieval process. Query specificity was annotated using our newly developed annotation method based on natural language questions and answers. We achieved reasonable performance for familiarity (Macro F1: 71.25\%) and query specificity results (Macro F1: 60.54\%). These findings suggest that eye-tracking features may offer insights into user needs and goals, which could help inform the design of more personalised and adaptive IR systems.

\begin{table}[t]
\centering
\caption{Effectiveness ($\mu\pm\sigma$) of binary classifiers for predicting query specificity using eye-tracking data.}
\adjustbox{max width=.48\textwidth}{
\sisetup{
    separate-uncertainty = true,
    table-align-uncertainty = true,
    detect-weight = true
}
\begin{tabular}{lS[table-format=2.2(2)]S[table-format=2.2(2)]S[table-format=2.2(2)]S[table-format=2.2]}  
\toprule
\textbf{Model} & \textbf{Accuracy} & \textbf{Macro F1} & \textbf{AUC} & \textbf{LOSO F1} \\ 
\midrule

Logistic Regression & 
70.09 \pm 2.59 & 
48.02 \pm 9.86 & 
70.91 \pm 6.37 & 
\cellcolor[HTML]{caebc0} \bfseries 58.11 \\ 

KNN & 
\cellcolor[HTML]{caebc0} \bfseries 71.24 \pm 6.44 & 
\cellcolor[HTML]{caebc0} \bfseries 60.54 \pm 12.70 & 
72.86 \pm 8.27 & 
53.34 \\

Gradient Boosting & 
69.90 \pm 5.03 & 
44.59 \pm 5.21 & 
\cellcolor[HTML]{caebc0} \bfseries 74.09 \pm 8.30 & 
44.17 \\ 

Random Forest & 
69.90 \pm 5.03 & 
44.59 \pm 5.21 & 
74.09 \pm 8.30 & 
40.31 \\

Decision Tree & 
69.90 \pm 5.03 & 
44.59 \pm 5.21 & 
74.09 \pm 8.30 & 
44.17 \\ 

\bottomrule
\end{tabular}
}
\label{table:specificity}
\end{table}

\myparagraph{Limitations and Future Work}

In our study, we find it difficult to automatically determine  query specificity. So far, we have only analysed the average model performance. Future work could explore features to better distinguish between topic familiarity and query specificity and conduct a more in-depth analysis of query specificity prediction results.
Another challenge in this study is the lack of data for models to generalise to unseen users. Additional data could be collected from participants to improve model generalisation. An expanded dataset would also enable exploring advanced models, such as deep learning models, extending beyond this study's five traditional machine learning models.

\begin{acks}
This research was conducted by the ARC Centre of Excellence for Automated Decision-Making and Society (ADM+S, \grantnum{ARC}{CE200100005}), and funded fully by the Australian Government through the \grantsponsor{ARC}{Australian Research Council}{https://www.arc.gov.au/}. We thank Kaixin Ji for providing access and support in using the dataset.
 This work was conducted on the unceded lands of the  Woi wurrung and Boon wurrung language groups of the eastern Kulin Nation. We pay our respect to Ancestors and Elders, past and present, and extend that respect to all Aboriginal and Torres Strait Islander peoples today and their connections to land, sea, sky, and community. 
\end{acks}

\balance
\bibliographystyle{ACM-Reference-Format}
\bibliography{00-refs}

\end{document}